\documentclass[twoside,leqno]{article}

\usepackage[letterpaper]{geometry}

\usepackage{siamproceedings}

\usepackage[T1]{fontenc}
\usepackage{amsfonts}
\usepackage{graphicx}
\usepackage{epstopdf}
\usepackage{enumitem}
\usepackage{algorithmic}
\ifpdf
  \DeclareGraphicsExtensions{.eps,.pdf,.png,.jpg}
\else
  \DeclareGraphicsExtensions{.eps}
\fi

\usepackage{tikz}
\usetikzlibrary{decorations.pathmorphing}
\usetikzlibrary{positioning}
\newsiamremark{remark}{Remark}
\newsiamremark{hypothesis}{Hypothesis}
\crefname{hypothesis}{Hypothesis}{Hypotheses}
\newsiamthm{claim}{Claim}

\usepackage{amsopn}

\newcommand\cM{{\mathcal M}}
\newcommand\cV{{\mathcal V}}
\newcommand\bC{{\mathbb C}}
\newcommand\bR{{\mathbb R}}
\newcommand\Tr{\mathrm{Tr}}
\newcommand\fhw{\mathfrak{hv}}
\newcommand\fgl{\mathfrak{gl}}
\begin{document}

\newcommand\relatedversion{}

\title{\Large The Categorical 't Hooft Expansion \relatedversion}
    \author{Davide Gaiotto\thanks{Perimeter Institute for Theoretical Physics
  (\email{dgaiotto@perimeterinstitute.ca}).}}

\date{}

\maketitle

% Copyright Statement
% When submitting your final paper to a SIAM proceedings, it is requested that you include
% the appropriate copyright in the footer of the paper.  The copyright added should be
% consistent with the copyright selected on the copyright form submitted with the paper.
% Please note that "20XX" should be changed to the year of the meeting.

% Default Copyright Statement
\fancyfoot[R]{\scriptsize{Copyright \textcopyright\ 2026 by SIAM\\
Unauthorized reproduction of this article is prohibited}}

% Depending on which copyright you agree to when you sign the copyright form, the copyright
% can be changed to one of the following after commenting out the default copyright statement
% above.

%\fancyfoot[R]{\scriptsize{Copyright \textcopyright\ 20XX\\
%Copyright for this paper is retained by authors}}

%\fancyfoot[R]{\scriptsize{Copyright \textcopyright\ 20XX\\
%Copyright retained by principal author's organization}}

%\pagenumbering{arabic}
%\setcounter{page}{1}%Leave this line commented out.

\begin{abstract} We review categorical aspects of 't Hooft's large $N$ expansion, which is expected to map any Quantum Field Theory of large matrices to a string theory. Our goal is to describe a general strategy to derive the string theory dual to given QFT, at least at the leading order in the 't Hooft expansion. The basic idea is to characterize the underlying worldsheet theory of the dual string theory as an extended 2d differential graded Topological Field Theory (dg-TFT), i.e. present an $A_\infty$-category of boundary conditions (``D-branes''). A basic aspect of the 't Hooft expansion is that D-branes arise from the addition of vector-valued degrees of freedom  to the QFT. We propose that formal deformations of such ``fundamental modifications'' must match the formal deformations of the dual D-branes, which in turn capture the $A_\infty$-category structure and thus the worldsheet dg-TFT. We discuss several systems for which a rigorous analysis along these lines is or should be possible.
\end{abstract}

\section{Introduction.}
The 't Hooft large $N$ expansion \cite{tHOOFT1974,tHooft:2002ufq,Witten:1979kh} is a general strategy to reorganize the perturbative expansion of a Quantum Field Theory (QFT) with $U(N)$ gauge group and $N \times N$-matrix valued fields.\footnote{The notion extends naturally to any non-exceptional gauge group.} See e.g. \cite{Marino:2004eq} for a review.
We can illustrate the 't Hooft expansion in the example of a matrix integral (aka ``Matrix Model'') \cite{Brezin:1978aa,Itzykson:1979fi,MIGDAL1983199,LandoZvonkin2004}:
\begin{equation}
	Z_{N}(\hbar,\lambda) \equiv \int dM \, e^{- \frac{1}{\hbar} \mathrm{Tr} \left(\frac12 M^2 + \frac{\lambda}{4} M^4\right)}
\end{equation}
Here $M$ is an $N \times N$ Hermitian matrix, $\hbar$ is a loop-counting parameter and $\lambda$ a coupling constant. The measure $dM$ is invariant under the adjoint action of $U(N)$ and normalized so that the integral is $1$ at $\lambda=0$. 

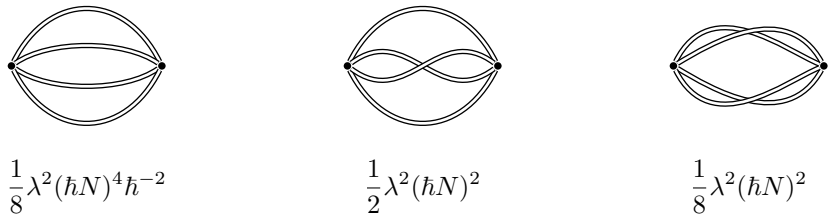
\begin{figure}[h]
\centering

\tikzset{
  prop/.style={double,double distance=1pt,line width=0.5pt}
}

%---------------- PLANAR ORBIT ----------------%
\begin{tikzpicture}[thick,scale=1,baseline=(current bounding box.center)]
  % two 4-valent vertices
  \node[fill,circle,inner sep=1pt] (v1) at (-1,0) {};
  \node[fill,circle,inner sep=1pt] (v2) at ( 1,0) {};

  % four parallel double-line propagators = planar ribbon graph
  \draw[prop] (v1) .. controls (-0.5, 1.0)  and (0.5, 1.0)  .. (v2);
  \draw[prop] (v1) .. controls (-0.5, 0.35) and (0.5, 0.35) .. (v2);
  \draw[prop] (v1) .. controls (-0.5,-0.35) and (0.5,-0.35) .. (v2);
  \draw[prop] (v1) .. controls (-0.5,-1.0)  and (0.5,-1.0)  .. (v2);

  % labels
  \node at (0,-1.6) {$\displaystyle \frac18 \lambda^2 (\hbar N)^4 \hbar^{-2}$};
\end{tikzpicture}
\hspace{2.0cm}
%---------------- NON-PLANAR ORBIT (SIZE 16) ----------------%
\begin{tikzpicture}[thick,scale=1,baseline=(current bounding box.center)]
  % two 4-valent vertices
  \node[fill,circle,inner sep=1pt] (w1) at (-1,0) {};
  \node[fill,circle,inner sep=1pt] (w2) at ( 1,0) {};

  % outer propagators (non-crossing)
  \draw[prop] (w1) .. controls (-0.5, 1.0)  and (0.5, 1.0)  .. (w2);
  \draw[prop] (w1) .. controls (-0.5,-1.0)  and (0.5,-1.0)  .. (w2);

  % inner pair: simple crossing, each ribbon intersects the other once
  \draw[prop] (w1) .. controls (-0.2, 0.6)  and (0.2,-0.6)  .. (w2);
  \draw[prop] (w1) .. controls (-0.2,-0.6)  and (0.2, 0.6)  .. (w2);

  % labels
  \node at (0,-1.6) {$\displaystyle \frac12 \lambda^2 (\hbar N)^2$};
\end{tikzpicture}
\hspace{2.0cm}
%---------------- NON-PLANAR ORBIT {1,2,3,4} ----------------%
\begin{tikzpicture}[thick,scale=1,baseline=(current bounding box.center)]
  % two 4-valent vertices
  \node[fill,circle,inner sep=1pt] (u1) at (-1,0) {};
  \node[fill,circle,inner sep=1pt] (u2) at ( 1,0) {};

  % four simple curved propagators, representative of orbit {1,2,3,4}
  \draw[prop] (u1) .. controls (-0.5, 1.0)  and (0.5, 0.2)  .. (u2);
  \draw[prop] (u1) .. controls (-0.5, 0.3)  and (0.5, 0.9)  .. (u2);
  \draw[prop] (u1) .. controls (-0.5,-0.3)  and (0.5,-0.9)  .. (u2);
  \draw[prop] (u1) .. controls (-0.5,-1.0)  and (0.5,-0.2)  .. (u2);

  % labels
  \node at (0,-1.6) {$\displaystyle \frac18 \lambda^2 (\hbar N)^2$};
\end{tikzpicture}

\caption{Some order $\lambda^2$ contributions to the perturbative expansion of the quartic matrix integral. Left: the leading (aka planar) contribution has $g=0$. Middle and right: the sub-leading contributions have $g=1$.} \label{fig:rib}
\end{figure}

A formal expansion of the integrand in powers of $\lambda$, followed by Gaussian integration, gives an asymptotic expansion for $Z_N(\hbar,\lambda)$ computed by Feynman diagrams. 
Each diagram has tetravalent vertices carrying a factor of $\lambda \hbar^{-1}$ and edges with a factor of $\hbar$. The combinatorics of matrix indices is represented graphically by promoting 
the edges to ribbons with oriented sides, joining in a planar way at the vertices. The sides of the ribbons form closed loops, each contributing a factor of $N$. See \cref{fig:rib}. We can denote the overall 
factor for a connected ribbon graph as 
\begin{equation}
	(\lambda \hbar^{-1})^V \hbar^E N^F = \lambda^V (N \hbar)^F \hbar^{-V+E-F} = \lambda^V (N \hbar)^F \hbar^{2g-2} \,, 
\end{equation} 
visualizing the ribbon graph as drawn on a Riemann surface with genus $g$. Adding up the Feynman diagram contributions with a fixed genus gives 
the coefficients $F_g(\lambda, N \hbar)$ of the 't Hooft expansion
\begin{equation}
	\log Z_N(\hbar, \lambda) \sim \sum_{g=0}^\infty \hbar^{2g-2} F_g(\lambda, N \hbar) \, ,
\end{equation}
which has the form of an $N \to \infty$  asymptotic expansion for $\log Z_N$ with fixed coupling $\lambda$ and 't Hooft coupling $N \hbar$. Note that the power series in $N \hbar$ converges to an actual function $F_g(\lambda, N \hbar)$ here \cite{BESSIS1980109}. This property is expected to be true in general.

The general expectation is that the 't Hooft expansion of any similar QFT coincides with the perturbative expansion of a String Theory \cite{Gross:1992tu,Aharony:1999ti,Kontsevich_AT1999}, i.e. the $F_g(\lambda, N \hbar)$ coefficients can be recast as 
\begin{equation}
	F_g(\lambda, N \hbar) = \int_{\cM_{g,0}} \omega_g(\lambda, N \hbar)
\end{equation}
for a collection of closed top forms $\omega_g(\lambda, N \hbar)$ on the moduli space $\cM_{g,0}$ of genus $g$ Riemann surfaces, computed as partition functions of 
a dual two-dimensional ``worldsheet'' QFT, which is topological in a cohomological sense (dg-TQFT).\footnote{Conventional String Theories are often presented in a ``conformal gauge'' in terms of a 2d Conformal Field Theory coupled to a ghost system and equipped with a BRST differential. The resulting 2d theory is topological in the cohomological sense, i.e. the stress tensor is BRST exact. The appearance of Riemann surfaces and $\cM_{g,0}$ is particularly natural in the conformal gauge, but can be justified without reference to it \cite{Kontsevich:1992ti,costello2006topologicalconformalfieldtheories}. } More generally, correlation functions of ``single-trace operators'' such as 
\begin{equation}
	\int dM e^{- \frac{1}{\hbar} \mathrm{Tr} \left(M^2 + \lambda M^4\right)} \prod_{i=1}^n \hbar^{-1} \mathrm{Tr} M^{k_i} 
\end{equation}
should arise from $n$-point correlation functions of local operators in the dg-TQFT, integrated over the moduli space $\cM_{g,n}$ of Riemann surfaces with $n$ punctures. See \cite{Gopakumar:2024jfq,Gopakumar:2011ev,Gopakumar:2003ns} and references therein for recent attempts to make this statement precise for Gaussian/free models.  

At the moment, there is no systematic strategy to recover the worldsheet dg-TQFT from the data of the original QFT. The dual dg-TQFT is (conjecturally) known for a collection of  examples, some of which form the backbone of holography: a powerful conjectural relation between QFT and quantum gravity \cite{Maldacena:1997re,Aharony:1999ti,David:1984tx,KAZAKOV1985282,Kontsevich:1992aa,Witten:1990hr,Gopakumar:1998ki,Dijkgraaf:2002fc,Aharony:2008ug}. Advances in our understanding of the 't Hooft expansion will have an important impact in both subjects. They may also facilitate a mathematical formalization and proof of examples of holography. 

The mathematical formalization of Topological Field Theory is well developed. In particular, it is well established that the data of a TFT (in a trivial Witt class) can be captured by the (higher) category of its boundary conditions \cite{lurie2009classificationtopologicalfieldtheories}. For a 2d dg-TFT, e.g. a TFT valued in complexes of vector spaces,  the data is an $A_\infty$-category which satisfies some extra dualizability axioms \cite{costello2006topologicalconformalfieldtheories,Kontsevich:1994dn,nadler2008constructiblesheavesfukayacategory,Barannikov:1997dwc,Kontsevich:2006jb,Herbst:2004jp}. We will not attempt to spell out these axioms here, partly because it is not yet established which general type of dg-TFT should appear in the context of the 't Hooft expansion. Outside that context, well known examples identified in the study of Homological Mirror Symmetry include the B-model dg-TFTs described by derived coherent sheaves on Calabi-Yau manifolds and A-model dg-TFTs described in general by Fukaya categories and sometimes as categories of D-modules or constructible sheaves. 

Curiously, this paradigm has not yet been applied to the 't Hooft expansion. This is our general goal: identify the categorical data of a worldsheet dg-TFT from calculations in the large $N$ QFT. 
Notice the use of a TFT vs TQFT notation above. The conventional definitions of string theory employ a local worldsheet 2d QFT which is topological in a cohomological sense: the theory is valued in complexes of vector spaces with differential given by the BRST charge, and the stress tensor is BRST exact. The primitive of the stress tensor can be used to define correlation functions which evaluate to closed top forms $\omega_{g,n}$ on $\cM_{g,n}$. We denote that as a dg-TQFT. We instead denote as a dg-TFT the result of a categorical definition. Given a dg-TQFT, it is usually possible to extract the data of a dg-TFT by a quasi-isomorphism which eliminates all states above an energy cutoff \cite{Witten:1982df}. Here we aim to bypass the dg-TQFT entirely and work directly with the dg-TFT data.

A certain amount of work is needed to convert the dg-TFT into the desired forms $\omega_{g,n}$ on $\cM_{g,n}$. This machinery is sometimes denoted as ``dg-TCFT''.
A further layer of complexity arises when integrating the $\omega_{g,n}$ forms on $\cM_{g,n}$, which is not compact. The necessary regularization procedures (``dealing with IR singularities in string theory'') are non-trivial and model-specific. They are an important part of the definition of a string theory. All of our calculations will involve genus $0$ contributions for which these challenges are manageable. 

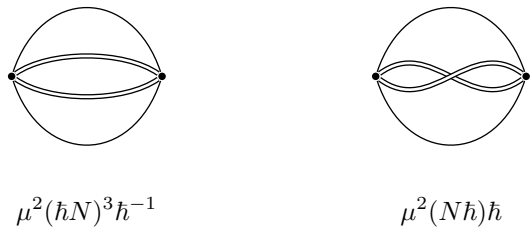
\begin{figure}[h]
\centering
\tikzset{
  prop/.style={double,double distance=1pt,line width=0.5pt}, % matrix ribbons
  boundary/.style={line width=0.5pt}                          % single-edge boundary
}

%---------------- DISK (g = 0, one boundary) ----------------%
\begin{tikzpicture}[thick,scale=1,baseline=(current bounding box.center)]
  % boundary vertices
  \node[fill,circle,inner sep=1pt] (v1) at (-1,0) {};
  \node[fill,circle,inner sep=1pt] (v2) at ( 1,0) {};

  % single-edge boundary loop, same style (width) as ribbon sides
  \draw[boundary] (v1) .. controls (-0.6, 1.2) and (0.6, 1.2) .. (v2);
  \draw[boundary] (v2) .. controls ( 0.6,-1.2) and (-0.6,-1.2) .. (v1);

  % two internal double-line propagators, planar (no crossing)
  \draw[prop] (v1) .. controls (-0.5, 0.35)  and (0.5, 0.35)  .. (v2);
  \draw[prop] (v1) .. controls (-0.5,-0.35)  and (0.5,-0.35)  .. (v2);

  % scaling
  \node at (0,-1.8) {$\displaystyle \mu^2 (\hbar N)^3 \hbar^{-1} $};
\end{tikzpicture}
\hspace{2.5cm}
%---------------- GENUS-1 WITH BOUNDARY (g = 1, one boundary) ----------------%
\begin{tikzpicture}[thick,scale=1,baseline=(current bounding box.center)]
  % boundary vertices
  \node[fill,circle,inner sep=1pt] (w1) at (-1,0) {};
  \node[fill,circle,inner sep=1pt] (w2) at ( 1,0) {};

  % single-edge boundary loop, same style (width) as ribbon sides
  \draw[boundary] (w1) .. controls (-0.6, 1.2) and (0.6, 1.2) .. (w2);
  \draw[boundary] (w2) .. controls ( 0.6,-1.2) and (-0.6,-1.2) .. (w1);

  % two internal double-line propagators, non-planar (crossing)
  \draw[prop] (w1) .. controls (-0.25, 0.55)  and (0.25,-0.55)  .. (w2);
  \draw[prop] (w1) .. controls (-0.25,-0.55)  and (0.25, 0.55)  .. (w2);

  % scaling
  \node at (0,-1.8) {$\displaystyle  \mu^2 (N\hbar) \hbar$};
\end{tikzpicture}

\caption{Some order $\mu^2$ contributions to the perturbative expansion of the quartic matrix integral with a fundamental modification. Left: the leading (aka planar) contribution has $g=0$, $b=1$. Right: the sub-leading contribution has $g=1$, $b=1$.} \label{fig:open}
\end{figure}

There is a well-understood strategy to add worldsheet boundaries to the 't Hooft expansion: modify the original QFT by adding ``vector valued'' fields transforming in the (anti)fundamental representation of $U(N)$ \cite{Witten:1979kh,KOSTOV1990181,Minahan:1991pv,Witten:1992fb,Gopakumar:1998ki, Karch:2002sh,Drukker:2005kx}. We will denote such manipulations as ``fundamental modifications''. For example, we could consider an integral 
\begin{equation}
	\widetilde Z_N(\hbar,\lambda,\mu) \equiv \int dM dv d\bar v\, e^{- \frac{1}{\hbar} \mathrm{Tr} \left(\frac12 M^2 + \frac{\lambda}{4} M^4\right)- \frac{1}{\hbar} \bar v (1+ \mu M^2) v} \, .
\end{equation}
If we formally expand in powers of $\lambda$ and $\mu$, the resulting Feynman diagrams will have new propagator edges from $v$-$\bar v$ contractions, forming closed loops which function as boundaries of the resulting Riemann surface. The resummed coefficients $\hbar^{2g-2+b} F_{g,b}(\lambda, \mu, N \hbar)$ with $b$ boundaries should match integrals over the moduli space $\cM_{g;b}$ of Riemann surfaces of genus $g$ with $b$ boundaries. See \cref{fig:open}.

The general expectation is that the integrand should be computed by the same dg-TQFT as before, but with boundaries. Different types of fundamental modifications are associated to different types of worldsheet boundaries, aka ``D-branes'', which are the objects in the $A_\infty$-category of boundary conditions we are after. We can invoke another general principle: the data of an $A_\infty$-category is essentially equivalent to the formal deformation theory of its objects \cite{Kontsevich:2000er}. Furthermore, only disk correlation functions should be needed for the calculation, i.e. the leading (planar) part of the genus expansion associated to the fundamental modifications.

The optimistic claim that all large $N$ systems which admit a 't Hooft expansion are dual to String Theories could thus be formulated as follows: 
\begin{claim}
	The planar formal deformation theory of fundamental modifications of a QFT which admits a 't Hooft expansion defines a 3d Calabi-Yau $A_\infty$-category. The full large $N$ expansion can be recovered from a string theory associated to the dg-TFT defined by the $A_\infty$-category.
\end{claim} 
At the moment, this program has only been partially realized in an handful of large $N$ systems. We will review a representative example in \cref{sec:chiral}. We devote \cref{sec:quantum} and \cref{sec:chern} to other collections of systems for which we foresee a rigorous mathematical analysis. 

The main message we want to convey is that the 't Hooft expansion and associated String Theories are ready to be mathematically formalized. After all, QFT perturbation theory is 
now rigorous \cite{costello2011renormalization}, so it should be possible to formalize the notion of fundamental modifications and their planar formal deformation theory as well, and prove that they define certain 2d dg-TFTs. Pushing the analysis beyond the planar order would give a perturbative formalization of an enormous class of string theory backgrounds. 

%To avoid confusion, we should point out an alternative notion of ``large $N$'' which leads to homological algebra constructions. The rank $N$ appears polynomially in individual terms in the 't Hooft expansion and in similar contexts. There are general strategies to ``analytically continue in $N$'' such calculations while dropping trace relations: work in a Deligne category \cite{Deligne2002CategoriesTensorielle} or replace $\fgl_N \to \fgl_{N+k|k}$ and take $k\to \infty$. They often play a role in mathematical work on holography \cite{Loday:1984aa,Costello:2017fbo,Zeng:2025pss}. See \cite{Zeng:2023lox} for an example with fundamental modifications. These steps may facilitate the definition of the 't Hooft expansion, but do not replace it. Once $N$ is treated as a continuous parameter, it is also possible to send it to $0$. This is essentially equivalent to 

\subsection{A Sharper Characterization.}
For completeness, we should mention a subtlety of the large $N$ expansion: 
%A useful feature of the example was the choice of a ``single trace'' action of the form $S = \Tr \cdots$, 
%and a ``meson'' action $\bar v S' v$ for the fundamental modification, which guarantees a simple 't Hooft reorganization of the expansion. In non-trivial QFT examples, renormalization 
%may require counter-terms which involve products of traces and/or mesons.  and require a careful treatment to preserve the existence of the 't Hooft expansion. A possible strategy is to promote the 
%couplings of singe-trace terms to auxiliary fields, which can be integrated out to produce more complicated terms. These subtleties are a counterpart of the ``IR divergences'' encountered on the string theory side.
the same QFT may admit multiple perturbative expansions and thus multiple distinct 't Hooft expansions. For example, consider the matrix integrals with the opposite sign for the $\Tr M^2$ term. One can define natural perturbative expansions around a minimum 
\begin{equation}
M^* = \operatorname{diag}\big(
\underbrace{\lambda^{1/2},\ldots,\lambda^{1/2}}_{N_{+}\ \text{times}},
\underbrace{-\lambda^{1/2},\ldots,-\lambda^{1/2}}_{N_{-}\ \text{times}}
\bigr) \,,
\end{equation}
of the action, with $N_\pm$ eigenvalues $\pm \lambda^{-\frac12}$, $N_+ + N_-=N$:
\begin{itemize}
	\item If $N_-=0$ (or $N_+=0$) the expansion has the same schematic form as before. 
	\item If $N_-=n$ is kept finite as we vary $N$ (or $N_+=n$), we can then break $M-M_*$ into: a matrix with $(N-n)^2$ elements which is treated as before, matrices with $n (N-n)$ elements which behave as the $v$, $\bar v$ of $n$ fundamental modifications, and a matrix of $n^2$ elements which behave as dynamical couplings for the fundamental modifications. The genus expansion now involves surfaces with boundaries and thus D-branes.
	\item If both $N_\pm$ scale as $N$, we get a model with multiple matrices of large size with two 't Hooft couplings $\hbar N_\pm$. This is dual to a two-parameter family of dual string theory backgrounds/worldsheet dg-TFTs.
\end{itemize}
The matter is further complicated by the fact that classical minima of the action receive quantum corrections, some of which are amplified by factors of $N$. 
\begin{definition}
We will refer to all the extra choices which go into the definition of the 't Hooft expansion for a QFT as the choice of a ``large $N$ saddle'' for the QFT. 
\end{definition}

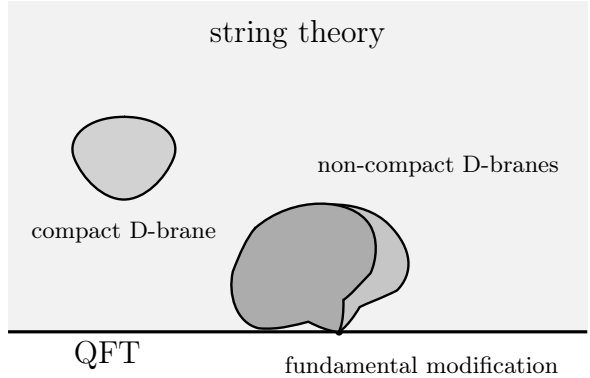
\begin{figure}[ht] 
\centering
\begin{tikzpicture}[scale=1.1, every node/.style={font=\small}]

  % Half-plane (string theory region)
  \fill[gray!10] (0,0) rectangle (7,4);
  \draw[very thick] (0,0) -- (7,0); % boundary

  % Labels for the bulk and boundary
  \node at (3.5,3.6) {\large string theory};
  % Move QFT left to avoid overlap
  \node[below] at (1.2,0) {\large QFT};

  % Common boundary point for both spikes
  \coordinate (FM) at (4,0);
  \fill (FM) circle (1.3pt);
  \node at (5.0,-0.4) {fundamental modification};

  % --- Compact D-brane blob in the bulk ---
  \begin{scope}[shift={(1.4,2.1)}]
    \fill[gray!35, draw=black, line width=0.9pt]
      (0,0.5)
      .. controls (0.45,0.5) and (0.7,0.25) .. (0.6,0)
      .. controls (0.5,-0.25) and (0.25,-0.5) .. (0,-0.5)
      .. controls (-0.25,-0.5) and (-0.5,-0.25) .. (-0.6,0)
      .. controls (-0.7,0.25) and (-0.45,0.5) .. cycle;
  \end{scope}

  \node at (1.4,1.2) {compact D-brane};

  % --- Non-compact D-brane 1 (symmetric top) ---
  \begin{scope}[shift={(4,0)}]
    \fill[gray!45, draw=black, line width=0.9pt]
      (-0.6,1.4)
      .. controls (-0.3,1.6) and (0.3,1.6) .. (0.6,1.3)
      .. controls (0.9,1.0) and (0.9,0.7) .. (0.7,0.5)
      .. controls (0.5,0.4) and (0.35,0.35) .. (0.25,0.3)
      % shared spike
      .. controls (0.18,0.18) and (0.10,0.08) .. (0.05,0.04)
      -- (0,0)
      -- (-0.05,0.04)
      .. controls (-0.10,0.08) and (-0.18,0.18) .. (-0.25,0.3)
      .. controls (-0.35,0.35) and (-0.5,0.4) .. (-0.7,0.5)
      .. controls (-0.9,0.7) and (-0.9,1.0) .. cycle;
  \end{scope}

  % --- Non-compact D-brane 2 (asymmetric top), same spike ---
  \begin{scope}[shift={(4,0)}, rotate=32]
    \fill[gray!65, draw=black, line width=0.9pt]
      (-0.2,1.6)
      .. controls (0.4,1.7) and (1.0,1.3) .. (1.0,0.9)
      .. controls (1.0,0.7) and (0.85,0.5) .. (0.7,0.4)
      .. controls (0.5,0.35) and (0.35,0.3) .. (0.25,0.3)
      % shared spike
      .. controls (0.18,0.18) and (0.10,0.08) .. (0.05,0.04)
      -- (0,0)
      -- (-0.05,0.04)
      .. controls (-0.10,0.08) and (-0.18,0.18) .. (-0.25,0.3)
      .. controls (-0.4,0.35) and (-0.6,0.4) .. (-0.8,0.55)
      .. controls (-1.0,0.75) and (-0.9,1.1) .. (-0.7,1.3)
      .. controls (-0.5,1.45) and (-0.35,1.55) .. cycle;
  \end{scope}

  % Label for the non-compact D-branes
  \node at (5.2,2.0) {non-compact D-branes};

\end{tikzpicture}

\caption{Schematic depiction of D-branes which can occur in a large $N$ expansion. The depiction refers to ``holographic'' examples of large $N$ dualities, where a $d$-dimensional QFT on a manifold $M_d$ is dual to a string theory with a 2d worldsheet theory of maps into a $(d+1)$-dimensional manifold with boundary $M_d$ (not to be confused with boundaries for the worldsheet itself).  Some D-branes can emerge from a choice of large $N$ saddle in the absence of a fundamental modification. We refer to these as ``compact'' D-branes: in a holographic setup they do not reach the $M_d$ boundary. Other D-branes emerge from explicit fundamental modifications of the QFT. The same modification may correspond to multiple ``non-compact'' D-branes depending on a choice of large $N$ saddle. In a holographic setup, the asymptotic shape near $M_d$ of the non-compact D-branes is determined by the choice of modification.}
\label{fig:holo-dbranes}
\end{figure}

The example illustrates several facts: 
\begin{itemize}
	\item The same QFT may admit multiple large $N$ saddles, which differ at the leading order $\hbar^{-2} F_0$. Each such saddle will correspond to a different worldsheet theory and associated $A_\infty$-category of boundary conditions.  
	\item The choice of large $N$ saddle may admit refinements which differ at the subleading order $\hbar^{-1} F_{0,1}$. Each such refinement will correspond to a boundary condition for the 
	worldsheet theory and an object in the associated $A_\infty$-category. We refer to such D-branes which emerge in the absence of a fundamental modification as ``compact''. 
	\item A fundamental modification also corrects a 't Hooft expansion at order $\hbar^{-1} F_{0,1}$. The same fundamental modification may lead to multiple corrected large $N$ saddles, 
	each giving an object in the $A_\infty$-category.  We refer to such D-branes as ``non-compact'', with an ``asymptotic shape'' determined by the choice of fundamental modification. 
\end{itemize}
These monikers are inspired by geometric examples of large $N$ dualities. See \cref{fig:holo-dbranes} for an illustration. It would be interesting to give a mathematical characterization of the different types of D-branes which can occur in a more general context.

In conclusion, the ``planar formal deformation theory of fundamental modifications'' of a QFT will depend on the choice of large $N$ saddle at order $\hbar^{-2} F_0$ and 
will include both deformations of the fundamental modifications and deformations of the choice of large $N$ saddle at order $\hbar^{-1} F_{0,1}$.
In the following Sections we will give examples of both contributions. 

Note also that the existence of D-branes makes the large $N$ expansion different from, say, the saddle evaluation of a conventional integral: large $N$ saddles which differ only in the compact D-brane content have the same leading $\exp \hbar^{-2} F_0$ exponent but different $\exp \hbar^{-1} F_{0;1}$ disk contributions. The latter contribution scales with the number of extra compact D-branes and may actually allow an interpolation between large $N$ saddles with different $F_0$, as we saw in the example where $N_-$ was either finite or of order $N$. 

The term ``large $N$ saddle'' refers to the expectation that the 't Hooft expansion is the asymptotic expansion which controls the actual large $N$ behaviour of $Z_N$. The very possibility to define multiple 't Hooft expansions raises the question of which of the expansions, if any, will actually control the actual large $N$ behaviour of $Z_N$. We will not address this important problem, which in holographic examples is tied to deep problems in quantum gravity. 

\section{Two-dimensional Chiral Gauge Theory} \label{sec:chiral}
The focus of this Section is an example which has a well-understood conjectural string theory dual \cite{Gopakumar:1998ki,Costello:2018zrm}: the B-model with target $SL(2,\bC)$. The associated worldsheet theory should thus admit a categorical description involving a derived category of coherent sheaves on $SL(2,\bC)$. The categorical 't Hooft analysis is analyzed in detail in \cite{Gaiotto:2024dwr}.
%An intrinsic definition of $\cV_N$ begins with a one-parameter family of chiral algebras $\cV(c)$ with small ${\cal N}=4$ super-Virasoro symmetry and strongly generated by an infinite tower of currents transforming into specific representations of the super-Virasoro algebra. The family is parameterized by a central charge $c$ and admits simple quotients $\cV_N$ when $c=3(N^2-1)$. 

The role of the ``QFT'' here is played by a family of chiral algebras (aka Vertex Operator Algebras) denoted as $\cV_N$, which capture a protected subsector of ${\cal N}=4$ SYM \cite{Beem:2013sza,Bonetti:2016nma,Li:2016gcb,Arakawa:2023cki,Gaberdiel:2025eaf,Bonetti:2025kan,Beem:2025guj}. The gauge theory definition of these chiral algebras 
involves an $N\times N$ (traceless) matrix-valued $\beta\gamma$ system and an analogous $bc$ system, equipped with a nilpotent BRST charge $Q$ of schematic form 
\begin{equation}
	Q = \oint  \Tr \left(\frac12 b [c, c] + c [\beta,\gamma] \right) \, ,
\end{equation}
whose cohomology\footnote{Mathematically, this is usually denoted as semi-infinite cohomology $H^{\frac{\infty}{2}+\bullet}\bigl(\widehat{\mathfrak{gl}}_N,\, \mathcal{S}(\mathfrak{gl}_N)\bigr)$, see  \cite{Feigin:1991aa}. The BRST charge is the VOA mode used as a differential.} is $\cV_N$.\footnote{Physically, the $bc$ system and BRST charge are the output of a gauge-fixing procedure for a 2d chiral $U(N)$ gauge field coupled to the $\beta \gamma$ system.} It is useful to assign to both $\beta$ and $\gamma$ scaling dimension $\frac12$ and to denote them collectively as $Z_a$, $a=1,2$. This makes an $SU(2)$ Kac-Moody symmetry of the model manifest, with BRST-closed currents $\Tr \, Z_a Z_b$. The chiral algebra also includes symmetrized traces
\begin{equation}
	\Tr \, Z_{(a_1} Z_{a_2} \cdots Z_{a_n)}
\end{equation}
of scaling dimensions $\frac{n}{2}$ as well as other single-trace operators with an extra derivative or factors of $b$ or $\partial c$.  

A natural target for the large $N$ expansion are correlation functions on $\bC P^1$. Given a choice of  BRST cohomology representatives, correlation functions are computed by Wick contractions for the free $Z_a$ and $bc$ systems and thus by a (finite) sum of Feynman diagrams, amenable of a 't Hooft reorganization. 
%An important simplifying feature is that the number of Wick contractions computing a given correlation function is finite, so the perturbative expansion truncates. 

Although our goal is to introduce fundamental modifications in the system and look at their formal deformations, it is instructive to set up a formal deformation theory for $\cV_N$ itself. 
We are interested in deformations preserving the structure of the underlying $U(N)$ gauge theory. We do so by considering deformations of the BRST differential: $Q \mapsto Q + \delta$. Focusing on $\bC P^1$ correlation functions, we require $\delta$ to be the linear combination of well-defined symmetry charges on $\bC P^1$ built from single-trace local operators:
\begin{equation}
	\delta_{[\alpha],n}  \equiv  \oint dz z^{-n-1} \Tr W_{[\alpha]}(z)\, , \qquad \qquad 0\leq n \leq 2 \Delta_\alpha-2 \, .
\end{equation}
Here $W_\alpha$ are the collection of all ``words'' built as a sequence of $Z_a$, $b$, $c$ or their derivatives. 
We denote by $[\alpha]$ a representative of the cyclic orbits of such words. We denote by $\Delta_\alpha$
the scaling dimension of $W_\alpha$. The range of $n$ is selected so that the modes are globally defined on $\bC P^1$. 

A deformation is admissible if it satisfies a BRST-anomaly cancellation condition:
\begin{equation}
	\{Q,\delta\} + \delta^2 = 0 \, .
\end{equation}
The expression on the left hand side contains terms linear in single-trace generators but also polynomial (``multi-trace'') terms. We will disregard the latter in order to define a ``planar'' Maurer-Cartan equation controlling formal perturbations.\footnote{The truncation step may seem a bit ad hoc. It would be better to set up a well-defined problem at all orders in $\hbar$ and then truncate it. A challenge is that we cannot hope to write a ``Maurer-Cartan equation valued in single-trace operators'': that would essentially be a classical equation of motion for the dual string theory, but string theory should be quantized with quantization parameter $\hbar^2$. In the context of a worldsheet description, the need for quantization arises when regularizing BRST anomalies from the boundary of ${\cal M}_{g,n}$ \cite{Bershadsky:1993ta,Bershadsky:1993cx,Eynard:2007hf,Fischler:1996ja,Kogan:1996zv}. Concretely, we may thus try to ``quantize'' the coefficients $C_\bullet$ of a general expansion
\begin{equation}
	\delta = \sum_{[\alpha],n} C_{[\alpha],n} \delta_{[\alpha],n} \, ,
\end{equation}
say by deforming the algebra of formal power series in the $C_\bullet$ coefficients to a non-commutative algebra or even an $A_\infty$ algebra, by an amount proportional to $\hbar^2$. It is then natural to write a deformed BRST anomaly condition \begin{equation}
	\{Q,\delta\} + (\delta,\delta) + (\delta, \delta, \delta) + \cdots = 0 \, .
\end{equation}
and systematically adjust the deformation to eliminate multi-trace terms from the expansion.  
It would be interesting to pursue further or improve this naive proposal and implement it at the level of correlation functions.}

The planar part of $\{Q,\delta\}$ defines a differential on the $\delta_{[\alpha],n}$ and the  planar part of $\delta^2$ defines a Lie algebra structure. The resulting dg-Lie algebra $\fhw$ is called the ``wedge'' algebra for the dg-chiral algebra. It coincides with the Lie algebra of divergence-free poly-vectorfields in $SL(2,\bC)$. For example, the zero modes of the Kac-Moody currents
\begin{equation}
	J_{(ab)} = \oint  dz\, \Tr Z_a Z_b(z)
\end{equation}
match the vector fields implementing the left action of $SL(2,\bC)$ on itself, while the modes 
\begin{equation}
	L_{-1} = \oint dz \,T(z)  \qquad \qquad L_{0} = \oint dz z \,T(z) \qquad \qquad L_{1} = \oint dz z^2 \,T(z) 
\end{equation}
of the stress tensor 
\begin{equation}
	T =  \Tr \left[ \frac12 \beta \partial \gamma - \frac12 \gamma \partial \beta + b \partial c \right]
\end{equation}
match the vector fields implementing the right action of $SL(2,\bC)$ on itself. 

Categorically, deformations of the large $N$ system should map to (rotation-invariant) deformations of the worldsheet dg-TFT, e.g. the cyclic cohomology of the category of boundary conditions. In a B-model with $SL(2,\bC)$ target, this indeed coincides with divergence-free poly-vectorfields in $SL(2,\bC)$. 

\subsection{Space-filling D-branes}
To access an interesting boundary condition for the worldsheet theory, we can add our first fundamental modification: an extra collection $I^A$, $J_A$ of (anti)fundamental $bc$ and $\beta\gamma$ system pairs with an overall $U(k|k)$ global symmetry. The $SU(N)$ gauge indices are omitted for clarity. This adds a term 
\begin{equation}
	Q^{(f)} = \oint dz \,I^A c \,J_A(z) \, ,
\end{equation}
to the BRST charge.\footnote{Mathematically, we are now working with $H^{\frac{\infty}{2}+\bullet}\bigl(\widehat{\mathfrak{gl}}_N,\, \mathcal{S}(\mathfrak{gl}_N \oplus \mathrm{Hom}(\bC^{k|k},\bC^{N} )\bigr)$.} 

Formal deformations of the fundamental modification now involve a $\delta^{(f)}$ symmetry generator built from 
\begin{equation}
	[\delta^{(f)}_{\alpha,n}]^A_B  \equiv  \oint dz z^{-n-1} I^A W_{\alpha} J_B(z) \, , \qquad \qquad 0\leq n \leq 2 \Delta_\alpha \, .
\end{equation}
The $U(k|k)$ index structure here is very important. We will come back to it momentarily.

A deformation is valid if 
\begin{equation}
	\{Q,\delta^{(f)}\} + \{Q^{(f)},\delta^{(f)}\} + \delta^{(f)}\delta^{(f)} = 0 \, .
\end{equation}
We will again focus on single-meson terms of the form $IWJ$, disregarding the rest as subleading. The $U(k|k)$ index structure works out so that the resulting ``mesonic wedge'' dg-Lie algebra can be written as $\fgl_{k|k}[B]$ for a dg-algebra $B$, which coincides with the algebra of holomorphic polynomial functions on $SL(2,\bC)$.

For example, the zero modes 
\begin{equation}
	M[1]^A_B \equiv \oint dz \,I^A J_B(z) \, ,
\end{equation}
generate $\fgl_{k|k}$ and correspond to the identity in $B$. The next non-trivial modes
\begin{equation}
	M[u_a]^A_B = \oint dz \,I^A Z_a J_B(z)  \, ,\qquad \qquad M[v_a]^A_B = \oint dz z \,I^A Z_a J_B(z)  \, ,
\end{equation}
correspond to the matrix elements ${u_a \choose v_a}$ of $SL(2,\bC)$ as elements of $B$. A simple calculation demonstrates that they commute and satisfy 
\begin{equation}
	u_1 v_2 - u_2 v_1 = \hbar N \, ,
\end{equation}
thus reconstructing the geometry of $SL(2,\bC)$. 

The way the $\fgl_{k|k}$ matrix structure promotes a Lie algebra to an algebra is completely analogous to the way formal deformations of boundary conditions in a dg-TFT are governed by an $A_\infty$ algebra rather than just an $L_\infty$ algebra: multiple copies of a boundary condition (aka Chan-Paton factors) can be deformed by matrix-valued boundary local operators and the associated Maurer-Cartan equation detects the full $A_\infty$ data. See \cite{Gaiotto:2024gii} for a recent discussion. We identify $B$ with the dg-algebra of (normalizable) local operators on the worldsheet boundary condition dual to this fundamental modification. It is clearly a B-model space-filling D-brane in $SL(2,\bC)$, i.e. the coherent sheaf ${\mathcal O}$.

As a check, one may observe the effect on $B$ of deformations $\delta$ of $Q$. This modifies the differential on $\delta^{(f)}$ by $\{\delta, \delta^{(f)}\}$ projected to single-meson terms and thus the dg-algebra $B$. Detailed calculations show a match with the deformations of the algebra of holomorphic functions on $SL(2,\bC)$
induced by divergence-free poly-vectorfields. Indeed, this gives a simple route for the identification of vector-fields: we can compute the commutator of $\delta_{[\alpha],n}$ modes and $M[u_a]$, $M[v_a]$ to see the action on $SL(2,\bC)$ coordinates.

\subsection{Local operators and non-holomorphic deformations}
In a BV formalism, the fields above are promoted to $(0,\bullet)$ forms by combining them with anti-fields. The BV action takes the form 
\begin{equation}
 S_{\mathrm{BV}} = \int_{\bC P^1} \Tr \left(b \bar \partial c + \beta \bar \partial \gamma + \frac12 b [c, c] + c [\beta,\gamma] \right) \, ,
\end{equation}
and the $\delta$ deformation appears as an extra interaction 
\begin{equation} \label{eq:smear}
	\sum_{[\alpha]} \int_{\bC P^1} \mu_{[\alpha]}(z) \, \Tr W_{[\alpha]}(z) \, ,
\end{equation}	 
for holomorphic sections $\mu_{[\alpha]}(z)$ of ${\cal O}(2 \Delta_\alpha-2)$. These are the deformations which 
preserve the structure of a holomorphic QFT. Although they appear as an ``interaction'' in the BV action  $S_{\mathrm{BV}}$,
it is easy to see that they do not actually change the action of the physical fields. Instead, they deform the BRST charge $Q$ by $\delta_{[\alpha],n}$ as discussed above. 

In perturbation theory, we can generalize the interaction to $\mu_{[\alpha]}$ being not holomorphic and valued in 
$(0,\bullet)$ forms. At the leading order, these deformations introduce a BRST anomaly proportional to $\bar \partial \mu_{[\alpha]}(z)$.
The $(0,1)$ form deformations do modify the action of the physical fields, so that the deformed partition function 
becomes a generating function of $\Tr W_{[\alpha]}$ correlation functions. This construction thus incorporates correlation functions in the 
formal deformation analysis.  

Notice that D-branes and morphisms in the B-model are typically defined and computed as Dolbeault complexes involving $(0, \bullet)$ forms on the target space. 
Although one can also work algebraically with \v{C}ech cohomology, the Dolbeault presentation allows one to introduce the BCOV/Kodaira-Spencer formalism for 
string theory calculations beyond the planar approximation \cite{Bershadsky:1993ta,Bershadsky:1993cx,Costello:2015xsa}. We do not have an analogous technology for 
algebraic presentations of the B-model. The dg-TFT which emerges from the enlarged deformation space of the BV action for $\cV_N$ has a hybrid geometric and algebraic presentation: 
the complex involves the Dolbeault operator along the ``holographic screen'' $\bC P^1$ direction but treats other directions in $SL(2,\bC)$ 
algebraically. It would be interesting to explore the implications of this observation beyond the planar approximation. 

There is also an interesting interplay between the formal deformations associated to the ``smeared'' chiral algebra local operators in \cref{eq:smear} and putative worldsheet local operators which appear in the description of chiral algebra correlation functions with $\Tr W_{[\alpha]}(z)$ insertions. In other examples of holography, e.g. \cite{Maldacena:2001km}, the world-sheet theory is ``non-compact'' and is expected to include both ``normalizable'' local operators which can be added to the world-sheet action as a formal deformation and ``non-normalizable'' local operators which cannot. 

\subsection{The role of ``backreaction''} \label{sec:back}
It is often useful to consider a limit where the 't Hooft coupling $N \hbar$ is small. At the leading order, this means neglecting all Feynman diagrams with loops, even 
these enhanced by factors of $N$. In this approximation, the action of $Q$ becomes a derivation, acting on individual fields and mapping them to the product of two fields. E.g. $c \mapsto c c$. This equips the space of fields with the structure of a (degree-shifted) co-algebra, dual to $\bC[\theta^1, \theta^2]$ for Grassmann odd $\theta^\alpha$. The free field OPE additionally provides a trace on the algebra, promoting it to a 2d Calabi-Yau algebra. This is the same algebra whose $A_\infty$ modules (i.e. complexes $d(z_1,z_2): V \to V$) describe D-branes for a B-model with $\bC^2$ target space. See \cite{Gaiotto:2024dwr} for details.

The whole 't Hooft expansion can be expressed in terms of the data of $\bC[\theta^1, \theta^2]$ and works equally well for any other compact 2d Calabi-Yau algebra $A$ which carry a $\bC^*$ action giving charge $2$ to the trace and satisfying a certain anomaly-cancellation condition. Indeed, we can associate to each such $A$ a free chiral algebra with fields $\Phi(z)$ valued in $A$ and BRST differential built from the Calabi-Yau structure, or better a BV action
\begin{equation}
	\int \frac12 (\Phi, \bar \partial \Phi) + \frac13(\Phi, \Phi, \Phi) \, .
\end{equation}
Viceversa, all chiral algebras defined by the BRST reduction of $\beta \gamma$ and $bc$ systems can be expressed in this form \cite{Li:2016gcb}. Nilpotency of the BRST differential at one loop gives an extra anomaly-cancellation condition. Such an algebra $A$ will define a non-commutative Calabi-Yau cone $X_{\mathrm{2d}}$ analogous to $\bC^2$. 

Recall that $SL(2,\bC)$ is related by a conifold transition to $O(-1) \oplus O(-1) \to \bC P^1$. A natural conjecture is that the 
outcome of the categorical 't Hooft expansion for the chiral algebra associated to $A$ will be a category of D-branes associated to a non-commutative 3d CY $X_{\mathrm{3d}}[\hbar N]$ obtained by a sort of conifold transition from $X_{\mathrm{2d}}(-1) \to \bC P^1$. It is easy to verify this conjecture in the $\hbar N \to 0$ limit. The planar BRST anomaly analysis of fundamental modifications gives a systematic algorithm to recover the analogue of 
the algebra of functions on $X_{\mathrm{3d}}[\hbar N]$ from a fundamental modification $I^A$, $J_A$.\footnote{The definition of the fundamental modification involves a choice of augmentation $A \to \bC$ selecting the ghost $c$ out of $\Phi$ entering $I^A c J_A$.} 

\subsection{Loop problems}
The planar categorical analysis sketched until now provides a candidate worldsheet dg-TFT for the string theory dual to the large $N$ chiral algebras. For the canonical example, we have a B-model with a geometric target $SL(2,\bC)$ and a possibly rigorous definition of the string theory genus expansion via Kodaira-Spencer/BCOV theory \cite{Bershadsky:1993cx,costello2006topologicalconformalfieldtheories}, assuming a generalization of the non-renormalization theorems from \cite{Costello:2015xsa} to include appropriate boundary correlation functions. The perturbative expansion of Kodaira-Spencer/BCOV theory should define a family of chiral algebras $\cV_{SL(2,\bC)}^{\hbar N}$ over power series in $\hbar$, parameterized by $\hbar N$. 

The individual $\cV_N$ are expected to be simple quotients of a continuous family $\cV^{\hbar N}$, defined e.g. with the help of a Deligne category \cite{Zeng:2025pss}. We have demonstrated that $\cV_{SL(2,\bC)}^{\hbar N}$ and $\cV^{\hbar N}$ agrees with it at the leading order in $\hbar$. Recent work \cite{Gaberdiel:2025eaf,Bonetti:2025kan} suggests that $\cV^{\hbar N}$ is unique. It thus must coincide with $\cV_{SL(2,\bC)}^{\hbar N}$! A rigorous proof of both statements would thus give a full proof of this large $N$ duality. 

For a more general $A$, we lack geometric intuition and an analogue to Kodaira-Spencer/BCOV theory. We would thus need to find novel ways to handle the divergences associated to non-compactness of $\cM_{g,n}$. A solution of the problem would probably hold important lessons about string theory in non-geometric backgrounds and other examples of holography. 

\subsection{Further generalizations}
We can define a more general class of chiral algebras which admit a large $N$ expansion by adding $SU(N)$ Kac-Moody currents to the mix. Gauged Kac-Moody currents can arise, for example, 
at 2d interfaces for 3d Chern-Simons gauge theory \cite{Witten:1988hf,Costello:2016nkh,Gaiotto:2017euk}. It is natural to scale the level $\kappa$ of the currents so that $N/(\kappa+N)$ is fixed and plays the role of a 't Hooft coupling.

The simplest example involves two copies of the Kac-Moody algebra with levels adding up to $-2N$. This describes 2d interfaces in 3d Chern-Simons gauge theory. After adding $bc$ ghosts, the BRST reduction results in a trivial chiral algebra. Nevertheless, fundamental modifications of the system are non-trivial. For example, adding a single set of fundamental chiral fermions (aka $N$ bc systems) leads to the famous collection of $W_N$ chiral algebras. Multiple collections of fermions of bosons lead to generalizations known as ``matrix-extended W-algebras'' \cite{Rapcak:2019wzw,Gaiotto:2023ynn}. The system has a conjectural holographic dual involving the A-model topological strings and ``coisotropic'' D-branes dual to the fundamental modifications \cite{GaiottoZhengInPrep}. 

It would be interesting to explore chiral algebras defined from more than two Kac-Moody factors or combining Kac-Moody currents and matrix-valued $\beta\gamma$ systems. It should involve an interesting combination of the B-model and A-model tools employed in better understood examples. 

\section{Double Howe duality and Holography} \label{sec:quantum}
At the very end of the previous Section we mentioned a system with a peculiar feature: the matrix-valued QFT itself is almost trivial, but still imbues the fundamental modifications with an interesting structure. The dual description must involve a string theory with an almost trivial closed strings sector which nevertheless supports a non-trivial category of D-branes. 
In this Section we will describe some instructive examples of such systems. 

\subsection{Pure Gauged Quantum Mechanics}
The simplest option is an $SU(N)$ gauged quantum mechanics without any matrix-valued 
matter fields. This system has a trivial space of states and no interesting observables. Mathematically, this can be understood as a quantum mechanics whose target space is the $SU(N)$ symplectic reduction of a point. The space of states is the trivial relative Chevalley complex $C^\bullet(\mathfrak{sl}_N|\mathfrak{sl}_N) = \bC$. 

As a fundamental modification, we consider $N$ 1d Dirac fermions, aka the system whose operator algebra is the complex Clifford algebra $\mathrm{Cl}(\bC^N)$:
\begin{align}
	\{ \psi^\alpha, \psi^\beta\} &= 0 \, , \cr
	\{\psi^\alpha, \bar \psi_\beta\} &= \hbar \delta^\alpha_\beta \, ,\cr
	\{ \bar \psi_\alpha, \bar \psi_\beta\} &= 0 \, ,
\end{align} 
with Greek letters denoting $SU(N)$ gauge indices, with $\psi^\alpha$ transforming in the fundamental representation and $\bar \psi_\alpha$ in the anti-fundamental. 
We will omit the gauge indices when possible, so e.g. $\bar \psi \psi \equiv \sum_\alpha \bar \psi_\alpha \psi^\alpha$ is an $SU(N)$ invariant ``meson'' operator. It generates a $U(1)$ 
flavour symmetry:
\begin{align}
	[\hbar^{-1} \bar \psi \psi,\psi^\alpha] &= - \psi^\alpha \, ,\cr
	[\hbar^{-1} \bar \psi \psi,\bar \psi_\alpha] &= \bar \psi_\alpha \, .
\end{align}

This system has a unique irreducible representation: the fermionic Fock space $\Lambda^\bullet \bC^N$. Coupling the system to the  $SU(N)$ gauged quantum mechanics implements a projection on $SU(N)$-invariant states and enforces an $SU(N)$ quantum Hamiltonian reduction of the Clifford algebra. There are two invariant states, $|0 \rangle$ and $|1\rangle$, respectively annihilated by all $\psi^\alpha$ or all $\bar \psi_\alpha$. Dual states $\langle 0|$ and $\langle 1|$ are respectively annihilated by all $\bar \psi_\alpha$ or all $\psi^\alpha$. 

We will include the choice of reference states in the past and future in the definition of a ``fundamental modification'' and corresponding ``D-brane'', as different choices lead to different 't Hooft expansions of correlation functions. It is natural to pair $\langle 0|$ and $|0 \rangle$ or $\langle 1|$ and $|1 \rangle$, as other pairings require the insertion of $N$ fermions and change the saddle analysis in a manner we will address later on. We denote the corresponding D-branes as $D_{1,0}$ and $D_{1,1}$. 

It is important to explore the formal deformations of multiple D-branes. We can thus consider a a collection of $m N$ 1d Dirac fermions, aka the system whose operator algebra is the complex Clifford algebra $\mathrm{Cl}(\bC^{mN})$:
\begin{align}
	\{ \psi_i^\alpha, \psi_j^\beta\} &= 0 \, ,\cr
	\{\psi_i^\alpha, \bar \psi^j_\beta\} &= \hbar \delta_i^j \delta^\alpha_\beta \, ,\cr
	\{ \bar \psi^i_\alpha, \bar \psi^j_\beta\} &= 0 \, ,
\end{align} 
with Greek letters denoting $SU(N)$ gauge indices and Latin letters denoting $U(m)$ flavour indices. We will omit the gauge indices when possible and flavour indices occasionally and use Einstein notation for the flavour indices unless otherwise noted.

The Hilbert space of the combined system is the $SU(N)$-invariant part of the Fock space $\Lambda^\bullet \bC^{mN}$. By skew Howe duality \cite{Weyl1939,Howe1989}, it can be presented as:
\begin{equation}
	H^{(N)}_m \equiv (\Lambda^\bullet \bC^{mN})^{SU(N)} = \bigoplus_{n=0}^m R^{U(m)}_{N,n}  \, ,
\end{equation}
where $R^{U(m)}_{N,n}$ denotes the irreducible representation of $U(m)$ labelled by a rectangular Young tableau with width $N$ and height $n$. We will refer to the integer $n$ as the ``baryon number''. The Fock space can also be presented as the tensor product of $m$ copies of $\Lambda^\bullet \bC^N$. Then $R^{U(m)}_{N,n}$ contains 
e.g. the state of the form $|1\rangle^{\otimes n} \otimes |0\rangle^{\otimes (m-n)}$ annihilated by the first $n$ flavours of $\bar \psi^i$ and the last $m-n$ flavours of $\psi_i$. 
The 't Hooft expansion of expectation values on such states is dual to a combination of  $n$ D-branes of type $D_{1,1}$ and $(m-n)$ of type $D_{1,0}$.

This choice admits a natural deformation: we can pick any $n$-dimensional subspace $V$ of $\bC^m$ and define a state $|V\rangle \in R^{U(m)}_{N,n}$ annihilated by all $v_i \bar \psi^i$ for $v \in V$ 
and all $\psi_i w^i$ for $w \in V^\perp$. So e.g. for $m=1$ we have $|0 \rangle = |\{0\}\rangle$, $|1 \rangle = |\bC\rangle$. For $m=2$, 
$|0 \rangle \otimes |0\rangle = |\{0\}\rangle$ and $|1 \rangle \otimes |1\rangle = |\bC^2\rangle$ are rigid, but $|0 \rangle \otimes |1\rangle$ is deformable to a $\bC P^1$ collection of states labelled by the possible lines in $\bC^2$.  We get to make this choice both in the past and in the future, with $\langle W|$ annihilated by all the $\psi_i w^i$ for $w \in W$ and all $v_i\bar \psi^i$ for $v \in W^\perp$. We have $\langle W|V\rangle=0$ if the dual spaces $W$ and $V$ are not in generic position, e.g. $V \cap W^\perp \neq {0}$, so we may want to impose $V \cap W^\perp = {0}$. 

More concretely, we can pick bases $v_i^a$ of $V$ and $w^i_a$ for $W$ and define
\begin{align}
	|V\rangle &= \prod_{a,\alpha} v_i^a \bar \psi^i_\alpha |0\rangle \, , \cr
	\langle W| &= \langle 0| \prod_{a,\alpha} \psi^\alpha_i w_a^i \, .
\end{align}
These definitions depend on the choices of bases only through an overall factor, so that $|V\rangle$ is a section of a line bundle over the Grassmanian $\mathrm{Gr}_{n,m}$: the $N$-th power of the determinant line. For example, 
\begin{equation}
	\langle W|V\rangle = (\det_{a,b} v^a w_b)^N \, .
\end{equation}
If $V$ and $W$ are in generic position, we can pick dual bases so that $v^a w_b = \delta^a_b$ and $\langle W|V\rangle =1$.

The resulting family of D-branes $D_{m,n}[V,W]$ is parameterized by an open subset of $\mathrm{Gr}_{n,m} \times \mathrm{Gr}_{n,m}$. There is a slightly better characterization of this space. 
The $SU(N)$-invariant meson operators 
\begin{equation}
	M^i_j = \hbar^{-1} \bar \psi^i_\alpha \psi_j^\alpha \, .
\end{equation}
 generate the action of the $U(m)$ flavour symmetry of the system. The symmetry is partly broken by the choice of reference states in the past and the future and $\hbar M$ acquires an expectation value which is easily computed as $\mu^i_{V, W;j} = N \hbar \sum_{a=1}^n w^i_a v^a_j$ when $v^a w_b = \delta^a_b$.

Notice the relation 
\begin{equation}
	M^i_j M^j_k = \hbar^{-2} \bar \psi^i_\alpha \psi_j^\alpha \bar \psi^j_\beta \psi_k^\beta = N\hbar^{-1} \bar \psi^i_\alpha \psi_k^\alpha  +m \hbar^{-1}\bar \psi^i_\alpha \psi_k^\alpha-\hbar^{-2}\bar \psi^i_\alpha \psi_k^\beta \bar \psi^j_\beta \psi_j^\alpha 
	= (N+m-n) M^i_k \, ,
\end{equation}
valid when acting on $SU(N)$-invariant states (aka after quantum Hamiltonian reduction) in $R^{U(m)}_{N,n}$, as well as $\Tr M = N n$. It is easy to see that correlation functions factorize at large $N$, so that the expectation value, say, of $M^i_j M^k_t$ is well approximated by the product of expectation values. The expectation value $\mu_{V, W}$ thus becomes a convenient label for the specific large $N$ saddle and is valued in the manifold $\mathcal{P}_{m,n}$ of $m \times m$ matrices which satisfy 
$\Tr \mu = n t$ and $\mu^2 = t \mu$ with $t = N \hbar$. This is a complex symplectic manifold which complexifies ${Gr}_{n,m}$ and deforms $T^* \mathrm{Gr}_{n,m}$. Any such matrix $\mu$ has rank 
exactly $n$ and can be factored as $\mu = \bar z z$ with $z \bar z = t$, defining $V$, $W$ in generic position. 
We can thus label the D-branes as $D_{m,n}[\mu]$ by points in $\mathcal{P}_{m,n}$.

We should mention a quantum version of this statement. One can use Howe duality to express each summand in the Hilbert space:
  \begin{equation}
	R^{U(m)}_{N,n}= (S^\bullet \bC^{m n})^{U(n),N}  \, ,
\end{equation}
as the subspace of a bosonic Fock space which transforms in the $\det^N$ representation of $U(n)$. In this presentation, we can introduce $2 n m$ bosonic oscillators 
\begin{align}
	[z^u_i,z^v_j] &= 0 \, , \cr
	[z^u_i,\bar z_v^j] &= \hbar \delta_v^u \delta^j_i \, ,\cr
	[\bar z_u^i,\bar z_v^j] &= 0 \, ,
\end{align}
and map the meson operators to 
\begin{equation}
	M^i_j = \hbar^{-1} \bar z^i_u z_j^u  \, .
\end{equation}
It is easy to verify that this $M$ satisfies the same linear and quadratic relations as above. 

This presentation maps the large $N$ expansion of the original system into a semiclassical limit \cite{Yaffe:1981vf}: the $SU(N)$-invariant quantum-mechanical system with Hilbert space $R^{U(m)}_{N,n}$ is well approximated by a classical mechanics system whose phase space is the Grassmanian $\mathrm{Gr}_{n,m}$, as a real locus $\bar z = z^\dagger$ in the complex symplectic phase space  $\mathcal{P}_{m,n}$. This justifies the above claims about the large $N$ expansion: correlation functions $\langle W| \cdots |V\rangle$ of products of mesons $M^i_j$ admit a semi-classical 
expansion around the vev $\mu$.\footnote{The $|V\rangle$ and $\langle W|$ states can be described as the charge $N$ part of coherent states for $z$ and $\bar z$ respectively.} 

In conclusion, we are led to seek a dual worldsheet dg-TFT which admits two boundary conditions $D_{1,0}$ and $D_{1,1}$, such that the direct sum of $m-n$ and $n$ such boundary conditions can be deformed to the collection of $D_{m,n}[\mu]$ boundary conditions parameterized by $\mathcal{P}_{m,n}$. We can access some information on morphisms between such D-branes from the tangent bundle to a point in $\mathcal{P}_{m+m',n+n'}$ which lies in the image of $\mathcal{P}_{m,n} \times \mathcal{P}_{m',n'}$ under the obvious embedding. 

The symplectic nature of the space of deformations is indicative of a 2d CY structure on this collection of D-branes. We will argue later on that the dual worldsheet dg-TFT should be understood as the product of an A-model with target $T^* \bR$ and an abstract 2d CY dg-TFT we will denote as $T_2$, which should include a collection of boundary conditions labelled by $\mathcal{P}_{m,n}$. Then $D_{m,n}[\mu]$ will be identified with the combination of an A-brane wrapping the $\bR$ base and a brane in $T_2$. 

We can propose two different looking descriptions for the brane category of $T_2$:
\begin{itemize}
	\item The quadratic equation satisfied by $\mu$ has obviously the appearance of a Maurer-Cartan equation for $m$ 	copies of a boundary condition. It corresponds to a dg-algebra with a ghost number $1$ element $C$ with non-trivial 	square $C^2 =CC$ and $d C = t C^2$. Then $\mu C$ is a Maurer-Cartan element and gives a moduli space  $\mathcal{P}_{m}$ of 	deformations which includes all the $\mathcal{P}_{m,n}$. This approach does not give insights on possible boundary local operators with other ghost numbers. 
	\item The bosonized description presents $\mathcal{P}_{m}$ as the space of representations of a Nakajma quiver with a single gauged node and a framing node of rank $m$. Intuitively, these should be thought of as bound states of $m$ ``non-compact'' flavour D-branes and any number $n$ ``compact'' D-branes. The parameter $t$ enters as a complex FI parameter, aka a linear superpotential term. 
\end{itemize}
The latter description as a category of quiver representations is our best candidate to describe the brane category of $T_2$.\footnote{One can reinforce the claim by recognizing the compact D-branes as a Grassmann parity-flipped version of the $N$ D-branes which could be used to engineer the  1d gauge theory.}

We will now consider another type of fundamental modification: 0d auxiliary variables $\theta^\alpha$ or $\bar \theta_\alpha$. These can be used in conjunction with the previous modifications to define ``baryon'' local operators such as 
\begin{equation}
	B(\bar s) = \prod_\alpha (\bar s^i \psi_i^\alpha) = \int d \bar \theta \exp \left[\bar s^i \psi_i^\alpha \bar \theta_\alpha\right] \, ,
\end{equation}
or an analogous 
\begin{equation}
	\bar B(s) = \prod_\alpha (\bar \psi^i_\alpha s_i ) = \int d \theta \exp \left[\theta^\alpha \bar \psi^i_\alpha  s^i\right] \, .
\end{equation}
These operators change the baryon number by one unit. A collection of correlation functions
\begin{equation}
	\langle W| \cdots B(s) \cdots |V\rangle \, ,
\end{equation}
with ellipsis denoting meson insertions, or generalizations with multiple $B$'s and $\bar B$'s, represents a ``composite''
fundamental modification whose large $N$ saddles should be dual to more complicated D-branes. Note that $|V\rangle$ can be created from $|0\rangle$ by acting with a sequence of $n$ $\bar B(s^a)$ operators, with $s^a$ spanning $V$, so without loss of generality we could consider expectation values on $|0\rangle$ of a collection of $\bar B(s^a)$ and $B(\bar s_b)$ 
in any sequence. 

In the bosonic description of the large $N$ limit, the baryon operators act as complex Lagrangian correspondences between $\mathcal{P}_{m,n}$ and $\mathcal{P}_{m,n\pm 1}$. 
It is easy to see that  
\begin{equation}
	B(\bar s) M^i_j - M^i_j B(\bar s)= \bar s^i s_j  \, ,
\end{equation}
for some $s_i$ such that $\bar s^i s_i = N$. We also have 
\begin{equation}
	\bar s^i M^j_i B(\bar s) =0 \, .
\end{equation}
Similarly, 
\begin{align}
	\bar B(s) M^i_j - M^i_j \bar B(s)= s_i \bar s^j  \, ,\cr
	\bar B(s) s_i M^i_j = 0 \, .
\end{align}
In terms of the $z$ and $\bar z$ variables, the correspondence can be equivalently defined by passing through the Hamiltonian reduction a simple correspondence between $\bC^{2(n+1)m}$ and $\bC^{2 n m}$: the co-normal to the locus $z^{n+1}_i = s_i$, $z^{a}_i = (z^a_i)'$ in $\bC^{nm+m} \times \bC^{nm}$. 

The large $N$ limit of a correlation function is thus associated to a concatenation of correspondences intersected with the Lagrangian submanifolds defined by $V$ and $W$. Each intersection point defines a large $N$ saddle and thus a D-brane. The expectation value $\mu$ of $\hbar M$ is constant in each interval, so the large $N$ saddle can be mapped to 
a sort of constructible sheaf of $T_2$ D-branes on $\bR$, with a locally constant choice of D-brane for $T_2$ which jumps in a specific way across the baryon insertions. Such constructible sheaves would naturally appear as a description of D-branes in the combined 3d dg-TFT with the $T^* \bR$ A-model factor. It would be interesting to develop this point further.

Notice that the deformation space of fundamental modifications includes the possibility of turning on 
a Hamiltonian built from meson operators. This will evolve the points in the phase space as we move along $\bR$. 
This option should be naturally included in the definition of a constructible sheaf. 

\subsection{Chern-Simons theory and Quantum Groups}
A notable variant of the above construction was recently analyzed in \cite{Gaiotto:2025nrd}: $SU(N)$ 3d Chern-Simons theory probed by 
fundamental modifications consisting of 1d fundamental Dirac fermions, i.e. Wilson lines carrying the $\Lambda^\bullet \bC^N$ representation. The CS theory was placed on $\bR \times S^2$ and the Wilson lines were placed along the $\bR$ factor at points in $S^2$ or slowly braided and merged to give rise to a link in $\bR \times S^2$.

Many aspects of the previous analysis apply to this system. In particular, the Hilbert space for $m$ parallel lines is a $q$-deformed version of $H^{(N)}_m$ and carries an analogous action of $U_q(\mathfrak{gl}_m)$, by quantum skew Howe duality \cite{Cautis_2014}. There also is a bosonized description, involving the quantization of a real slice of a twisted $GL(n)$ character variety
deforming $P_{m,n}$. Remarkably, the character variety describes a space of D-branes for the A-model with target $M_t$, the deformed $A_1$ singularity. This plays the same role as $T_2$ before. 

Given a choice of large $N$ saddle for the correlation function of a knotted Wilson loop in $\bR \times S^2$,
the evaluation of the $U_q(\mathfrak{gl}_m)$ generators at each $S^2$ slice gives again a sort of constructible sheaf, now 
valued in A-model D-branes in $M_t$. This is interpreted as an A-brane in $T^* \bR \times M_t$. More generally, the A-model with  
$T^* \bR \times M_t$ target space is the expected large $N$ dual to 3d Chern-Simons theory on $\bR \times S^2$. 

The 3d Chern-Simons setting is very rich and still only partially explored. We devote the next Section to it. 

\subsection{2d BF theory and the Yangian}
Another interesting almost-trivial large $N$ theory is 2d $SU(N)$ BF theory, which can be understood mathematically as the B-model with target $\mathrm{pt}/SL(N)$ \cite{Losev:2017zgw}. 
The elementary fields are a 2d connection and a Lagrange multiplier $B$ enforcing flatness of the 2d connection. 

One-dimensional fundamental modifications of the theory gave one of the first examples of ``Twisted Holography'' \cite{Ishtiaque:2018str}. Namely, we can couple 
the 2d gauge fields to $m$ flavours of 1d fundamental fermions to define an interface in 2d. The algebra $A_{m}^{(N)}$ of local operators on the interface can be identified with the $SL(N)$ quantum Hamiltonian reduction of 
\begin{equation}
	\mathrm{Cl}(\bC^{m N}) \otimes U(\mathfrak{sl}_N) \otimes U(\mathfrak{sl}_N) \, .
\end{equation}
If we denote the $U(\mathfrak{sl}_N)$ generators as $B_\pm$, generators for this algebra include central elements $\Tr B_+^n$ and $\Tr B_-^n$ and mesons
$\bar \psi^i B_\pm^n \psi_j$. It is not difficult to verify that the latter generate the image of the Yangian algebra $Y_\hbar(\mathfrak{gl_m})$ \cite{Dedushenko:2020yzd,Moosavian:2021ibw}.\footnote{The references consider bosonic fundamental oscillators. The analysis with fermions is analogous up to some signs.} More precisely, the generating functions
\begin{equation}
	T_\pm[z] = \bar \psi \frac{1}{z \mp B_\pm} \psi
\end{equation} 
are transfer matrices in the (anti)fundamental representation for an RTT presentation of the Yangian. The quantum determinants of $T_\pm[z]$ can be expressed in terms of $\Tr B_\pm^n$.
A ``vacuum'' of the 2d BF theory will specify values for these central elements, and should be prescribed for a well-defined large $N$ limit. 

The large $N$ limit again becomes a semi-classical limit for the algebra of meson operators. The semi-classical limit of the Yangian leads to a phase space which can be described as a space of
holomorphic-topological connections in $\bR \times \bC$. The connections may have ``Dirac'' singularities. We expect the location and charge of these singularities to be controlled by 
the expectation values of $\Tr B_\pm^n$, but we will not attempt to make the statement precise here. In a canonical vacuum with $\Tr B_\pm^n\sim 0$ we expect some singularity of magnitude $\hbar N$, analogous to the twist of the character variety found in the Chern-Simons setup. Holomorphic-topological connections are naturally interpreted as  D-branes for the combination of an A-model with target $T^* \bR$ and a B-model with target $\bC$. The Dirac singularity should represent a back-reaction modifying this naive AB-model geometry to some other 2d CY setting $T_2^{\mathrm{BF}}$.

If we evolve in time the configuration of 1d Wilson lines, we will again get some kind of constructible sheaf on $T^* \bR$ valued in $T_2^{\mathrm{BF}}$ D-branes. The large $N$ dual is plausibly identified with the combination of an A-model on $T^* \bR$ and $T_2^{\mathrm{BF}}$.

\subsection{General 1d systems} 
The data of a compact 2d CY algebra $A$ can be also used to build a BV action for a 1d gauged quantum mechanics with adjoint fields valued in $A$. 
\begin{equation}
	\int_\bR \frac12 (\Phi, d \Phi) + \frac13(\Phi, \Phi, \Phi) \, .
\end{equation}
Before back-reaction, the natural setting for this is the combination of an A-model on $T^* \bR$ and the dg-TFT $T_2^{0}[A]$ associated to $A$. 
We expect holographic dual geometries which can be interpreted as back-reacted deformations of such a dg-TFT.

A categorical analysis should follow a similar route as the chiral algebra setup, with the
free OPE replaced by the Moyal product and a BRST differential $[(\Phi, \Phi, \Phi),\cdot]$. Adding as a fundamental modification the familiar set of $2mN$ fermionic oscillators 
and looking at planar commutation relations, we can obtain a collection of phase spaces ${\cal P}_{m,n}[A]$, possibly infinite-dimensional, describing D-branes in a 2d CY category $T_2^{\hbar N}[A]$.
Then the full category should combine the A-model in $T^* \bR$ and $T_2^{\hbar N}[A]$ as before.

The case of $A=\bC[\theta_1,\theta_2]$ is particularly interesting. The associated dg-TFT is a B-model on $\bC^2$.
The un-modified system is a topological (aka zero Hamiltonian) version of the celebrated matrix Quantum Mechanics: the two ghost number $0$ fields are a matrix $X$ and its conjugate momentum $P$. Famously, $X$ can be diagonalized to reduce the operator algebra to a symmetric product of $N$ copies of a Weyl algebra $[x^i,p_i]=0$. 

Fundamental modifications of this system were studied to great effect, in a different guise, in \cite{Costello:2017fbo}. See also \cite{Gaiotto:2020dsq, Gaiotto:2023ynn}. This extends older work \cite{Gross:1990md,Polychronakos:2006nz,Gaiotto:2005gd,Betzios:2017yms} on the ``non-singlet sectors'' of the 
matrix Quantum Mechanics. The literature focusses on the case of $m$ bosonic modifications, i.e. the $SU(N)$ quantum Hamiltonian reduction of 
the combination of a matrix-valued Weyl algebra and $m$ vector-valued Weyl algebras. The resulting algebra admits a map from the shifted affine Yangian of $\mathfrak{gl}_m$.
Here we are most interested in the case of $m$ fermionic modifications, i.e. we combine the matrix-valued Weyl algebra with the Clifford algebra with $2 m N$ generators. 
A map from the shifted affine Yangian of $\mathfrak{gl}_m$ can be built in the same manner. The generators are the mesons
\begin{equation}
	e^i_{m,n;j} = \bar \psi^i :X^n P^m: \psi_j \, ,
\end{equation}
where the colons indicate a symmetrization over all possible relative orders of the $X$ and $P$ matrices. 

We can again treat the large $N$ limit as a semi-classical limit. In a sector of baryon number $n$ we will obtain an (infinite-dimensional) phase space $P_{m,n}^{N \hbar}(\bC^2)$. We expect these phase spaces to describe a deformation of the category of D-branes for the B-model on $\bC^2$. It would be interesting to compute the deformation explicitly and generalize it to other choices of $A$. 

The setup can be generalized to interfaces in 2d BF theory by adding copies of  $U(\mathfrak{sl}_N)$. It would be interesting to explore this generalization as well.

\section{More on 3d Chern-Simons theory} \label{sec:chern}
The large $N$ limit of 3d Chern-Simons theory with $U(N)$ gauge group is still only partially understood. If the theory is placed on $S^3$, 
the proposed large $N$ dual is the A-model with target the resolved conifold \cite{Gopakumar:1998ki,Ooguri:1999bv,Ooguri:2002gx}. Some generalizations of the proposal involve discrete quotients of $S^3$ \cite{Halmagyi:2003mm,Marino:2009dp} and show rich patterns of large $N$ saddles.
 If the theory is placed on $\bR \times S^2$, a recent proposal reviewed above 
suggests the large $N$ dual is the A-model with target $T^* \bR \times M_t$ \cite{Gaiotto:2025nrd}. For a general three-manifold $M_3$, the large $N$ dual does not appear to be understood. 
We will now discuss how the question could be formulated mathematically with the help of the categorical 't Hooft expansion.

The perturbative expansion of 3d Chern-Simons theory on a manifold $M_3$ is mathematically well-understood \cite{Axelrod:1991vq,Axelrod:1993wr}. Formally, we have a sum over Feynman integrals with a propagator $P(x,y)$ which is a two-form on $M_3 \times M_3$ and satisfies 
\begin{equation}
	d P(x,y) = \delta^{(3)}(x,y) \, ,
\end{equation}
as well as a gauge-fixing condition. The Feynman integrals for a Feynman diagram $\Gamma$ with $V$ vertices involve an integral over $M_3^V$ of a product of propagators
associated to edges. The integrands are singular, but the singularity can be handled either by a judicious real blow-up of $M_3^V$ along diagonals or by regularizing the propagators,
say by passing to a Schwinger time representation of $P$ in terms of the heat kernel and cutting off small Schwinger times. 

To probe the system, we can add $m$ 3d topological fundamental degrees of freedom. This adds 
\begin{equation}
	\int u^i d_A v_i + a_i^j u^i v_j 
\end{equation}
terms to the action, with $A$ being the $U(N)$ gauge field (promoted to a full form in a BV formalism) and $u^i$ and $v_i$ being form-valued (anti) fundamental fields. We included 
a coupling to a $U(m)$ connection $a^i_j$ which will play an important role in our analysis (and possibly promoted $u$ and $v$ to sections of a non-trivial $U(m)$ bundle). 

Classically, this action is gauge-invariant (better, satisfies the BV master equation) iff $a$ is flat, i.e. $d a + [a,a]=0$. In a BV formalism, these constraints arise from the 
BV bracket of the interaction with itself giving $(d a + [a,a])^j_i u^i v_j$. The propagator of the $u,v$ fields will depend on $a$: 
\begin{equation}
	d_a P_{uv}(x,y) = \delta^{(3)}(x,y) \, .
\end{equation}
Quantum-mechanically, couplings such as $a$ can be renormalized. The BV master equation/BRST anomaly constraint can also be deformed. We expect it will indeed be deformed in this situation,
with extra contributions proportional to $u^i v_j$. The anomaly itself is local and should only be a functional of $a$ itself and the local choice of gauge-fixed propagator. The simplest non-trivial contribution should have the schematic form 
\begin{equation}
	N \hbar \int_{|y-x|=\epsilon} P_{uv}(x,y) P(x,y) \, .
\end{equation}
 
 Accordingly, the classical constraint that $a$ is a flat connection on $M_3$ will be modified by planar contributions.
Each solution of the deformed flatness conditions should define a consistent fundamental modification and thus a D-brane in the dual theory. 
Better, the BV formalism will naturally lead to a formal deformation theory and thus to an $A_\infty$-category of ``deformed flat connections on $M_3$''. 
Flat connections on $M_3$ are naturally interpreted as D-branes for an A-model with $T^* M_3$ target space. The deformed notion of flat connection 
thus gives implicitly a deformation of the A-model with $T^* M_3$ target space. 

It would be interesting to compute this deformation and compare it with the general expectation for the large $N$ dual of the CS theory: an A-model on some version of $(T^*-\{0\})M_3$
with a symplectic form which has period $N \hbar$ on the unit sphere in the fiber at each point. 

\subsection{Physical problems}
Many other fundamental modifications are available in 3d Chern-Simons theory. Some have direct physical interest. For example, 
one could couple $U(N)$ Chern-Simons theory to physical matter fields with kinetic terms which depend explicitly on a metric in $M_3$. 
This fundamental modification should be dual to some yet-unknown type of A-model D-brane. It would be very nice to identify it. 

The authors \cite{Aharony:2019suq} considered an intermediate situation: 3d Holomorphic-Topological matter fields \cite{Aganagic:2017tvx}. The action is similar as for the topological case, 
but the holomorphic derivatives are dropped from the kinetic term and $a$ is now a holomorphic-topological connection. 
The dual D-brane is expected to be a ``coisotropic'' D-brane in the A-model \cite{Kapustin:2001ij}. It would be interesting to study the large $N$
expansion of the corresponding Holomorphic-Topological factorization algebra \cite{Budzik:2022mpd,Wang:2024tjf}.

\section*{Acknowledgments.}
We would like to thank D. Ben-Zvi, K. Budzik, K. Costello, J. Kulp, A. Lopez, S. Sanjurjo and Y. Soibelman for useful discussions and careful comments on the draft. This research was supported in part by a grant from the Krembil Foundation. Research at Perimeter Institute is supported in part by the Government of Canada through the Department of Innovation, Science and Economic
Development Canada and by the Province of Ontario through the Ministry of Colleges and Universities.

% SIAM recommends using BibTeX
% if using BibTeX
\bibliographystyle{siamplain}
\bibliography{proceedings}
\end{document}